\documentclass{jpsj3}
\usepackage{txfonts}
\usepackage{cases}
\usepackage{amsmath,amssymb} 
\usepackage[usenames]{color}
\newcommand{\vect}[1]{\mbox{\boldmath $#1$}}
\newcommand{\svect}[1]{\mbox{\boldmath {\scriptsize${#1}$}}}

\title{Mean-Field Theory Is Exact For the Random-Field Model \\with Long-Range Interactions}

\author{Junichi Tsuda\thanks{E-mail: tsuda@stat.phys.titech.ac.jp} and Hidetoshi Nishimori}
\inst{Department of Physics, Tokyo Institute of Technology, Megro-ku, Tokyo 152-8551, Japan} 

\abst{We study the classical spin model in random fields with long-range interactions  
and show the exactness of the mean-field theory under certain mild conditions. 
This is a generalization of the result of Mori for the non-random and spin-glass cases. 
To treat random fields, we evoke the self-averaging property of a function of random fields, 
without recourse to the replica method. 
The result is that the mean-field theory gives the exact expression of the canonical free energy for systems with power-decaying interactions 
if the power is smaller than or equal to the spatial dimension.}

\begin{document}
\maketitle

\section{Introduction}\label{sc:intro}
Long-range interacting systems attract attention in recent years since such systems have some peculiar properties, for example, negative specific heat in the microcanonical ensemble and ensemble inequivalence. \cite{A235}$^-$\cite{rev}
The latter property is that the physical properties differ between microcanonical ensemble and canonical ensemble.
A typical example of long-range interaction is the power-law potential which decays as $1/r^\alpha$, where $r$ is the distance between particles/sites.
When the effective range of interaction is almost the same as the system size, the system is non-additive: The whole system cannot be divided into subsystems under the condition that the energy of the whole system is equal to the sum of the energies of subsystems, resulting in the peculiar properties.
In astrophysics, long-range interacting systems have been investigated by many researchers, for example, for  self-gravitating systems. \cite{A235,HT71,138}

An important property of some of the long-range interacting systems is the exactness of the mean-field theory, which is defined that the free energy is exactly equal to that of the corresponding mean-field system. 
Many studies on specific examples suggest that long-range interacting systems have this property. \cite{B61}$^-$\cite{JPA36}
This property has been proved to hold for a class of generic non-random spin systems and spin-glasses by Mori.\cite{E84,E82,Mori}

However, there has been no report so far on the exactness of the mean-field theory of long-range interacting spin systems in random fields.\cite{SR_comment}
Random fields give each site characteristics different from other sites whereas long-range interactions tend to erase strong site dependence due to their averaging properties over many sites.
This implies that long-range interactions and random fields have conflicting effects, which may be worth detailed studies.
It should also be noticed that the corresponding mean-field model has been reported to have peculiar properties.\cite{MFS}
Another point to be remarked is that Mori used the replica method with integer replica number to discuss the spin-glass case, which makes his analysis incomplete. It is thus worthwhile to further study the effects of randomness in long-range interacting systems.
These observations motivate us to investigate the present system.

Our basic strategy is to generalize the method of Mori to accommodate random fields without recourse to the replica method.
A long-range interacting system in random fields is introduced in Section \ref{sc:model}.
In Section \ref{sc:main} we calculate the free energy and then prove the exactness of the mean-field theory
for systems without conservation of magnetization. 
Also, conditions are given for the exactness of the mean-field theory for systems with conserved magnetization.
Section \ref{sc:sum} summarizes this paper.

\section{Model}\label{sc:model}
In this section, a random-field spin model with long-range interactions is introduced. 
The system size is $N=L^d$, where $L$ is the linear size and $d$ is the spatial dimension.
Periodic boundary conditions are imposed. 
The Hamiltonian is defined as
\begin{equation}
{\mathcal{H}}=-\frac{J}{2}\sum_{i,j=1,i\neq j}^{N}{K(\vect{r} _{i,j})\,S_{i}\,S_{j}}-\sum_{i=1}^{N}{h_{i}\,S_{i}}\label{eqn:HD}\qquad(J>0),
\end{equation}
where $S_i$ is a general classical spin variable at site $i$ with a bounded value, for example, an  Ising spin $S_i=\pm 1$, $K(\vect{r})$ represents the interaction potential whose range is long in the sense as defined below, and $h_i$ denotes the random field at site $i$. 
The variable $\vect{r} _{i,j}$ is for the relative position of site $i$ and site $j$.
The distribution of random fields is arbitrary as long as it satisfies a mild condition to be specified later.

This paper deals with the potential as introduced by Mori\cite{E84,E82,Mori} defined through a non-negative function defined in $\mathbb{R}$ satisfying $\phi(\vect{r})=\phi(-\vect{r})$, 
\begin{equation}
K(\vect{r})=\gamma^d\phi(\gamma\vect{r})\ge0\label{eqn:dfH},
\end{equation}
where $\gamma$ is a positive number.
The function $\phi(\vect{r})$ is supposed to satisfy the following conditions,
\begin{equation}
|\phi(\vect{r})|<\psi(r),\quad |\nabla\phi(\vect{r})|<\frac{\mathit{d}}{\mathit{d}r}\psi(r)\quad(\forall \vect{r},|\vect{r}|=r),
\end{equation}
where $\psi(\vect{r})$ is twice-differentiable, convex, integrable, and defined in $(0,\infty)$.

The parameter $\gamma$ corresponds to the inverse of the interaction range because \vect{r} appears as the combination $\gamma\vect{r}$ in eq. (\ref{eqn:dfH}). 
To consider long-range potentials, we take the limit $\gamma\rightarrow0$ in two ways:
(i) The non-additive limit, $\gamma\rightarrow0$ with $\gamma L=1$, and (ii) the van der Waals limit,
$L\rightarrow\infty$ first and then $\gamma\rightarrow0$. 
In the non-additive limit, the range of interaction is comparable with the linear size of the system, $\gamma^{-1}\propto L$, and the system is non-additive. 
A typical example is the power-law potential $K(\vect{r})\propto1/r^\alpha\,(0\le\alpha<d)$, in which case $\phi(\vect{r})\propto 1/r^\alpha$.\
In the van der Waals limit, the interaction range is long but is much smaller than the system size, and the system is additive

The potential $K(\vect{r})$ is supposed to be normalized,
\begin{equation}
\label{eqn:normK}
		\sum_{i=1}^{N}K(\vect{r}_i)=1.
\end{equation}
In this paper, the exactness of the mean-field theory means that the free energy of the model (\ref{eqn:HD}) is exactly equal to that of the mean-field model (infinite-range model) ,
\begin{equation}
\label{eqn:HMF}
{\mathcal{H}}_{\mathrm{MF}}=-\frac{J}{2N}\sum_{i,j=1}^{N}{S_i\,S_j}-\sum_{i}{h_i\,S_i}.
\end{equation}

\section{Exactness of the Mean-Field Theory}\label{sc:main}
We now prove the exactness of the mean-field theory.
\subsection{Variational Expression of the Free Energy}
This section first introduces coarse-grained variables to replace the microscopic spin variables.
The whole system is divided into many subsystems $\{B_p\}_{p=1}^{\Omega}$, each of which is of size $l^d$, and $\Omega=(L/l)^d$ is the number of subsystems.
The linear length $l$ of a subsystem is much larger than the lattice spacing, which is taken to be unity for simplicity, and is much smaller than $L$, $1\ll l \ll L$. 
The center site $\vect{r} _p$ of a subsystem $B_p$ is chosen to represent the location of the subsystem.

Let us define a coarse-grained variable $m_p$ of subsystem $B_p$ as 
\begin{equation}
\label{eqn:DefMp}
m_p=\frac{1}{l^d}\sum_{i\in B_p}S_i.
\end{equation}
We take the continuum limit, $L\rightarrow\infty$ and $l\rightarrow\infty$ with $L/l \rightarrow\infty$. 
Under this limit, the location \vect{x} $(\in {\mathbb{I}}^d=[0,1]^d)$ in $B_p$ is defined as $\vect{x}=\vect{r}_p/L$ and the coarse-grained variable as $m(\vect{x})=m_p$. 
The interaction potential should be normalized under the continuum limit, and hence it is defined as
\begin{equation}
\label{eqn:DefU}
U(\vect{x})=\lim_{L\rightarrow\infty}L^dK(L\vect{x}),
\end{equation}
with the normalization of the potential (\ref{eqn:normK}) being modified as
\begin{equation}
\label{eqn:normU}
\int_{{\mathbb{I}}^d}\mathit{d}^dx\, U(\vect{x})=1,
\end{equation}
where $\mathbb{I}$ means the interval from $0$ to $1$ $(\mathbb{I}=[0,1])$.
The free energy per spin for fixed magnetization $m$ is expressed in terms of the coarse-grained variables
as long as we consider a potential in the van der Waals limit or a power-law potential $r^{-\alpha}$ with
$0\le \alpha <d$, as described in Appendix\ref{app:1}.
The result is
\begin{align}
f(\beta,m)=\lim_{L,l,L/l \rightarrow \infty}{\left(-\frac{1}{\beta L^d}\ln{Z(\beta,m)}\right)}=\inf_{m(\svect{x})}\left\{\mathcal{F}\{\beta;m(\vect{x})\},\int_{\mathbb{I}^d}\mathit{d}^dx\,m(\vect{x})=m\right\}\label{eqn:RLf},
\end{align}
where $\beta$ is the inverse temperature. The generalized free energy per spin ${\mathcal{F}}\{\beta;m(\vect{x})\}$ is expressed as
\begin{multline}
{\mathcal{F}}\{\beta;m(\vect{x})\}=-\frac{J}{2}\iint_{{\mathbb{I}}^d \bigotimes {\mathbb{I}}^d}{\mathit{d}^dx\, \mathit{d}^dy\, U(\vect{x}-\vect{y})m(\vect{x})m(\vect{y})}\\
+J\int_{{\mathbb{I}}^d}{\mathit{d}^dx\,\mathcal{T}_{\beta}^{-1}\{m(\vect{x})\}m(\vect{x})}-\frac{1}{\beta}\int_{{\mathbb{I}}^d}{\mathit{d}^dx\left[\ln{Z_0\{\beta(J\,\mathcal{T}_{\beta}^{-1}\{m(\vect{x})\}+h)}\}\right]}\label{eqn:RLGF},
\end{multline}
where $[\cdots]$ represents the configurational average over the distribution of random fields $\{h_i\}$,
and the function $\mathcal{T}_{\beta}^{-1}\{\cdot\}$ is the inverse of $\mathcal{T}_{\beta}\{\cdot\}$ defined as 
\begin{equation}
\frac{1}{\beta}\frac{\partial}{\partial s}\left[\ln{Z_0}\{\beta(Js+h)\}\right]=\mathcal{T}_{\beta}\{s\}.
\label{config_average}
\end{equation}
The quantity $Z_0$ is the trace of the exponential $e^{tS}$ over the single spin variable $S$,
\begin{equation}
Z_0(t)=\mathrm{Tr}\,\mathit{e}^{tS}.\label{eqn:z0}
\end{equation}
The distribution of random fields is supposed to yield finite values of the average and variance 
around the saddle point discussed in Appendix A for 
quantities in the symbol of configurational average, e.g. $\ln{Z_0\{\beta(J\,\mathcal{T}_{\beta}^{-1}\{m(\vect{x})\}+h)}\}$ in eq. (\ref{eqn:RLGF}).
Similarly, the free energy per spin of the mean-field model is given as
\begin{equation}
		f_{\mathrm{MF}}\{\beta,m\}=-\frac{J}{2}m^2+J\mathcal{T}_{\beta}^{-1}\{m\}m-\frac{1}{\beta}\left[\ln{Z_0\{\beta(J\,\mathcal{T}_{\beta}^{-1}\{m\}+h)}\}\right]\label{eqn:MFf}.
\end{equation}

\subsection{Bounds for the Free Energy}
In this section, we derive inequalities on the free energy for fixed magnetization. Using the derived inequality, we prove the exactness of the mean-field theory in the next section.

Since the mean-field free energy (\ref{eqn:MFf}) is equal to the generalized free energy (\ref{eqn:RLf}) with $m(\vect{x})=m~(\forall\vect{x})$, the following inequality holds trivially,
 \begin{align}
	f(\beta,m)\leq\mathcal{F}\{\beta;m(\vect{x})=m\}=f_{\mathrm{MF}}\{\beta,m\}.\label{inq:UL}
\end{align}

We next derive a lower bound for the generalized free energy $\mathcal{F}\{\beta,m(\vect{x})\}$.
If we define the Fourier coefficient of the potential as 
\begin{equation}
U_{\svect{n}}=\int_{{\mathbb{I}}^d}\mathit{d}^dx\,  U(\vect{x})\cos(2\pi\vect{n}\cdot\vect{x})\quad (\forall \vect{n}\in\mathbb{Z}^d),
\end{equation}
the following inequality holds
\begin{equation}
U_{\svect{n}}\leq\int_{{\mathbb{I}}^d}\mathit{d}^dx\,  U(\vect{x})=1.\label{eqn:condition}
\end{equation}
Under periodic boundary conditions, the local magnetization can also be expanded into a Fourier series 
\begin{equation}
m_{\svect{n}}=\int_{{\mathbb{I}}^d}\mathit{d}^dx\, m(\vect{x})\exp(2\pi i\vect{n}\cdot\vect{x}).
\end{equation}
 With the Fourier expression, the first term in the generalized free energy (\ref{eqn:RLGF}) is rewritten as
\begin{align}
\iint_{{\mathbb{I}}^d\bigotimes{\mathbb{I}}^d}\mathit{d}^dx\,\mathit{d}^dy\, U(\vect{x}-\vect{y})m(\vect{x})m(\vect{y})=\sum_{\svect{n}}U_{\svect{n}}|m_{\svect{n}}|^2.\label{eqn:u1}
\end{align}
In the same way, we can obtain the following expression to be used later,
\begin{align}
\int_{{\mathbb{I}}^d}\mathit{d}^dx\, m(\vect{x})^2=\sum_{\svect{n}}|m_{\svect{n}}|^2.\label{eqn:u2}
\end{align}

Let us denote the largest coefficient of $U_{\svect{n}}$ with $\vect{n}\neq\vect{0}$ as $U_{\mathrm{max}}(\neq0)$.The first term in the generalized free energy (\ref{eqn:RLGF}) can then be upper bounded as
\begin{align}
\iint_{{\mathbb{I}}^d\bigotimes{\mathbb{I}}^d}\mathit{d}^dx\,\mathit{d}^dy\, U(\vect{x}-\vect{y})m(\vect{x})m(\vect{y})&\leq m^2+U_{\mathrm{max}}\sum_{\svect{n}\neq\svect{0}}|m_{\svect{n}}|^2\notag\\
&=m^2+U_{\mathrm{max}}\left\{\int_{{\mathbb{I}}^d}\mathit{d}^dx\, m(\vect{x})^2-m^2\right\}\label{ineq:1},
\end{align}
where $m$ stands for $m_0$.
With this inequality, as shown in Appendix \ref{app:GFE}, the generalized free energy is lower-bounded as
\begin{equation}
{\mathcal{F}}\{\beta;m(\vect{x})\}\geq
f_{\mathrm{MF}}\{\beta,m\}+U_{\mathrm{max}}\left\{\int_{{\mathbb{I}}^d}\mathit{d}^dx\,f_{\mathrm{MF}}^{*}\{\beta^{*},m(\vect{x})\}-f_{\mathrm{MF}}^{*}\{\beta^{*},m\}\right\}\label{eqn:1},
\end{equation}
where the function $f^{*}_{\mathrm{MF}}(\beta^{*},m)$ is defined as 
\begin{align}
f^{*}_{\mathrm{MF}}\{\beta^{*},m\}&=-\frac{J}{2}m^2+J\frac{\mathcal{T}_{\beta}^{-1}\{m\}}{U_{\mathrm{max}}}m-\frac{1}{\beta U_{\mathrm{max}}}\left[\ln{Z_0\left\{\beta U_{\mathrm{max}}\left(J\,\frac{\mathcal{T}_{\beta}^{-1}\{m\}}{U_{\mathrm{max}}}+\frac{h}{U_{\mathrm{max}}}\right)\right\}}\right]\notag\\
&=-\frac{J}{2}m^2+J(\mathcal{T}^{*}_{\beta^{*}})^{-1}\{m\}m-\frac{1}{\beta^{*}}\left[\ln{Z_0\left\{\beta^{*}\left(J\,(\mathcal{T}^{*}_{\beta^{*}})^{-1}\{m\}+\frac{h}{U_{\mathrm{max}}}\right)\right\}}\right]
\end{align}
with $\beta^{*}=\beta U_{\mathrm{max}}$.
We have defined the function $\mathcal{T}^{*}_{\beta^*}\{\cdot\}$ as 
\begin{equation}
\mathcal{T}^{*}_{\beta^{*}}\{s\}=\frac{1}{\beta^{*}}\frac{\partial}{\partial s}\left[\ln{Z_0}\{\beta^{*}(Js+h/U_{\mathrm{max}})\}\right].
\end{equation}
The quantity $f^{*}_{\mathrm{MF}}\{\beta^{*},m\}$ can be understood as the free energy of the mean-field model
 at inverse temperature $\beta^{*}$ with
the magnitude of the random field being greater by constant factor $1/U_{\mathrm{max}}$.

\begin{figure}
\centering
\includegraphics[width=0.45\columnwidth]{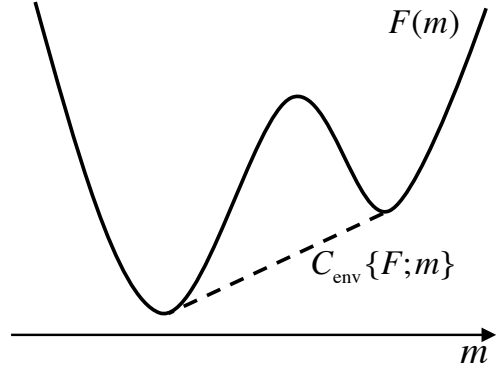}
\caption{The function $F(m)$ in bold and its convex envelope $C_{\mathrm{env}}\{F;m\}$ as dashed.
Where these two coincide, only the former is drawn.}
\label{fig:envelope}
\end{figure}

As seen in eq. (\ref{eqn:RLf}), the free energy is the lower limit of the generalized free energy, and therefore eq. (\ref{eqn:1}) leads to 
\begin{align}
f(\beta,m)\geq f_{\mathrm{MF}}\{\beta,m\}+U_{\mathrm{max}}\left\{\inf_{m(\svect{x})}\left\{\int_{{\mathbb{I}}^d}\mathit{d}^dx\,f_{\mathrm{MF}}^{*}\{\beta^{*},m(\vect{x})\},\int_{\mathbb{I}^d}\mathit{d}^dx\,m(\vect{x})=m\right\}-f_{\mathrm{MF}}^{*}\{\beta^{*},m\}\right\}.\label{eqn:LRL}
\end{align}
By the way, the following equation is known to hold,\cite{E84}
\begin{equation}
\inf_{m(\svect{x})}\left\{\int_{{\mathbb{I}}^d}\mathit{d}^dx\,f_{\mathrm{MF}}^{*}\{\beta^{*},m(\vect{x})\},\int_{\mathbb{I}^d}\mathit{d}^dx\,m(\vect{x})=m\right\}
=\mathrm{C_{env}}\big{\{}f^{*}_{\mathrm{MF}}\{\beta^{*},m\}\big{\}},\label{inq:LL}
\end{equation}
where $C_{\mathrm{env}}\{F;m\}$ is the convex envelope of a function $F(m)$. See Fig. \ref{fig:envelope}.

With eqs. (\ref{inq:UL}), (\ref{eqn:LRL}) and (\ref{inq:LL}), we obtain
\begin{equation}
f_{\mathrm{MF}}\{\beta,m\}-U_{\mathrm{max}}\Delta f^{*}_{\mathrm{MF}}\{\beta^{*},m\}\le f\{\beta,m\}\le f_{\mathrm{MF}}\{\beta,m\},\label{ineq}
\end{equation}
where the $\Delta f^{*}_{\mathrm{MF}}\{\beta^{*},m\}$ is defined as
\begin{equation}
\Delta f^{*}_{\mathrm{MF}}\{\beta^{*},m\}=f^{*}_{\mathrm{MF}}\{\beta^{*},m\}-\mathrm{C_{env}}\big{\{}f^{*}_{\mathrm{MF}}\{\beta^{*},m\}\big{\}}.
\end{equation}

\subsection{Exactness of the Mean-Field Theory (I): Non-Conserved Magnetization}
Let us first analyze the simple case of non-conserved magnetization.
Since the Fourier coefficient $U_{\mathrm{max}}$ is equal to or smaller than $1$,
eq. (\ref{ineq:1}) can be simplified as
\begin{align}
\iint_{{\mathbb{I}}^d\bigotimes{\mathbb{I}}^d}\mathit{d}^dx\,\mathit{d}^dy\, U(\vect{x}-\vect{y})m(\vect{x})m(\vect{y})\leq \sum_{\svect{n}}|m_{\svect{n}}|^2
=\int_{{\mathbb{I}}^d}\mathit{d}^dx\, m(\vect{x})^2.
\end{align}
It is not difficult to verify that the above inequality reduces eq. (\ref{eqn:1}) to
\begin{align}
{\mathcal{F}}\{\beta;m(\vect{x})\}\geq\int_{{\mathbb{I}}^d}\mathit{d}^dx\,f_{\mathrm{MF}}\{\beta,m(\vect{x})\}\label{eqn:3}.
\end{align}
Since the free energy of non-conserved system $f(\beta)$ is the minimum of ${\mathcal{F}}\{\beta;m(\vect{x})\}$
and that of the mean-field theory $ f_{\mathrm{MF}}(\beta)$ is the minimum of the above right-hand side, we find
\begin{align}
f(\beta)\geq f_{\mathrm{MF}}(\beta).
\end{align}
It also follows from eq. (\ref{inq:UL}) that $f_{\mathrm{MF}}(\beta)$ is lower-bounded by $f(\beta)$,
$ f_{\mathrm{MF}}(\beta)\ge f(\beta )$.
We therefore conclude the exactness of the mean-field theory,
\begin{align}
f(\beta)= f_{\mathrm{MF}}(\beta).
\end{align}

\subsection{Exactness of the Mean-Field Theory (II): Conserved Magnetization}

Equation (\ref{ineq}) indicates that the mean-field theory is exact for systems with conserved magnetization, $f\{\beta,m\}=f_{\mathrm{MF}}\{\beta,m\}$, when the mean-field free energy modified by $U_{\rm max}$ is convex, $\Delta f_{\mathrm{MF}}^{*}\{\beta^{*},m\}=0$. 
In particular, a potential in the van der Waals limit has $U(\vect{x})=\delta (\vect{x})$ as shown in Appendix \ref{app:vdW}, which means $U_{\svect{n}}=1$ for any $\vect{n}$ including $U_{\rm max}=1$.
It follows that $\beta^{*}=\beta$ and $f^{*}_{\mathrm{MF}}\{\beta^{*},m\}=f_{\mathrm{MF}}\{\beta, m\}$.
We therefore find that the mean-field theory is exact when $\Delta f_{\mathrm{MF}}\{\beta,m\}=0$
for potentials in the van der Waals limit.

For a general potential, the condition of convexity $\Delta f_{\mathrm{MF}}^{*}\{\beta^{*},m\}=0$ is clearly
unsatisfied when the second derivative is negative,
\begin{equation}
\frac{\partial^2}{\partial m^2} f_{\mathrm{MF}}^{*}\{\beta^{*},m\}<0
\end{equation}
as seen in Fig.  \ref{fig:envelope}. 
This fact can also be verified through the relations
\begin{equation}
\left.\frac{\partial^2\mathcal{F}\{\beta\,;\,m(\vect{x})\}}{\partial m_{-\svect{n}'}\partial m_{\svect{n}}}\right|_{m(\svect{x})=m}
=\left(-JU_{\svect{n}}+J\frac{\partial \mathcal{T}_{\beta}^{-1}\{m\}}{\partial m}\right)\delta_{\svect{n}'\svect{n}}\label{eqn:2d}
\end{equation}
and
\begin{align}
\frac{\partial^2}{\partial m^2}f^{*}_{\mathrm{MF}}\{\beta^{*},m\}=\frac{1}{U_{\mathrm{max}}}\left(-JU_{\mathrm{max}}+J\frac{\partial \mathcal{T}_{\beta}^{-1}\{m\}}{\partial m}\right)\label{eqn:3d}.
\end{align}
See Appendix \ref{app:GFE} for a derivation of eq. (\ref{eqn:2d}).
These equations indicate that the generalized free energy $\mathcal{F}\{\beta\,;\,m(\vect{x})\}$ does not
take its minimum if the second derivative of $f^{*}_{\mathrm{MF}}\{\beta^{*},m\}$ is negative
because eq. (\ref{eqn:2d}) becomes negative at $U_{\svect{n}}=U_{\mathrm{max}}$.

When the second derivative of  $f^{*}_{\mathrm{MF}}\{\beta^{*},m\}$ is positive and
$\Delta f^{*}_{\mathrm{MF}}\{\beta^{*},m\}\ne 0$, it is not possible to draw a definite conclusion
on the exactness of the mean-field theory.

\section{Summary}\label{sc:sum}
We have shown the exactness of the mean-field theory for spin systems with long-range interactions.
When the magnetization is not conserved, the mean-field theory is exact as long as the interaction
potential is in the van der Waals limit or the power of the potential $\alpha$ satisfies
$0\le \alpha <d$.
For systems with conserved magnetization, the mean-field theory is exact for a range of magnetization
where the modified mean-field free energy is convex.

These results generalize those of Mori who derived similar conclusions for systems without randomness and for spin-glass cases
using the replica method with integer replica number.
An advantage of our approach is that we did not use the mathematically ambiguous replica method
to treat randomness.
It is an important future problem to develop a method to discuss the spin-glass case
with long-range interactions without using replicas.

\begin{acknowledgment}
We thank Dr. Takashi Mori for kind discussions.
This work was supported by JSPS KAKENHI Grant number 23540440.

\end{acknowledgment}

\appendix

\section{Variational Expression of the Free Energy}\label{app:1}
This section derives the variational expression of the free energy in eq. (\ref{eqn:RLGF}).
Following closely the method of Mori,\cite{E84,E82,Mori} we can show that the long-range interaction term in the Hamiltonian
is expressed with the coarse-grained variable $m_p$ as 
\begin{equation}
\mathcal{H}=\overline{\mathcal{H}}+Ng_1(L,l,\{S_i\}),
\end{equation}
where
\begin{equation}
\overline{\mathcal{H}}=-\frac{J}{2}\sum_{p,q=1}^{\Omega} { U_{pq} m_p m_q}- \sum_{i}{h_i S_i} \label{eqn:CGHD}.
\end{equation}
The potential $U_{pq}$ in eq. (\ref{eqn:CGHD}) is defined as 
\begin{equation}
U_{pq}=\sum_{i\in B_p} \sum_{j\in B_q}K(\mbox{\boldmath{$r$}}_i-\mbox{\boldmath{$r$}}_j),
\end{equation}
and the function $g_1(L,l,\{S_i\})$ converges to zero in the limit $L\rightarrow \infty,
l\rightarrow \infty$ as long as we consider the van der Waals limit or the power potential $r^{-\alpha}$
with $0\le \alpha <d$.

Let us rewrite the partition function with fixed magnetization $m$ in terms of eq. (\ref{eqn:CGHD}),
\begin{align}
Z(\beta,m)&={\mathrm{Tr}} \exp{(-\beta \mathcal{H})}\delta\left(\frac{\sum_{i=1}^{N}S_i}{N}-m\right)\notag\\
&={\mathrm{Tr}} \idotsint_{\mathbb{R}^\Omega} \prod_{p}\mathit{d}m_p\,\delta \left (l^d m_p-\sum_{i\in B_p}{S_i} \right)\delta\left(\frac{\sum_{t=1}^{\Omega}m_t}{\Omega}-m\right)\notag\\&\times\exp{\beta\left( \frac{J}{2}\sum_{q,r} { U_{qr} m_q m_r}+ \sum_{i}{h_i S_i}+L^d g_1(L,l,\{S_i\})\right)}\label{eqn:PF1}
\end{align}
with the inverse temperature $\beta$.
The Fourier-transformed expression of the delta function reduces the partition function to 
\begin{align}
Z(\beta,m)
&=\frac{1}{(2\pi)^\Omega}{\mathrm{Tr}} \idotsint_{\mathbb{R}^\Omega \bigotimes (\mathit{i}\mathbb{R})^\Omega} \prod_{p}\mathit{d}m_p\mathit{d}(\beta J\, \overline{m}_p)\delta\left(\frac{\sum_{t=1}^{\Omega}m_t}{\Omega}-m\right)\notag\\
&\times\exp{\beta\left( \frac{J}{2}\sum_{q,r} { U_{qr} m_q m_r}-J\sum_{q}l^d \overline{m}_qm_q+\sum_{q}\sum_{i\in B_q}{( J\,\overline{m}_q+h_i)\,S_i}+L^d g_1(L,l,\{S_i\})\right)}.\label{eqn:PF2}
\end{align}
With the definition (\ref{eqn:z0}), the trace in eq. (\ref{eqn:PF2}) is rewritten as
\begin{align}
Z(\beta,m)&=\left(\frac{\beta J}{2\pi}\right)^\Omega\idotsint_{\mathbb{R}^\Omega \bigotimes (\mathit{i}\mathbb{R})^\Omega} \prod_{p}\mathit{d}m_p\mathit{d}\overline{m}_p\delta\left(\frac{\sum_{t=1}^{\Omega}m_t}{\Omega}-m\right)\notag\\
&\times\exp{\left( \frac{\beta J}{2}\sum_{q,r} { U_{qr} m_q m_r}-\beta J\sum_{q}l^d \overline{m}_qm_q+\sum_{q}\sum_{i\in B_q}\ln{Z_0\{\beta( J \,\overline{m}_q+h_i)}\}+\beta L^d g_1(L,l)\right)}\label{eqn:PF}.
\end{align}
The function $g_1(L,l)$ in eq. (\ref{eqn:PF}) does not depend on $\{S_i\}$ and tends to zero in the limit
$L\rightarrow\infty, l\rightarrow\infty$ for the following reason. The partition function (\ref{eqn:PF1}) can be bounded using the maximum $g_1^{\mathrm{max}}(L,l)$ and the minimum $g_1^{\mathrm{min}}(L,l)$ of $g_1(L,l,\{S_i\})$ among all configuration of spins $\{S_i\}$.
Then the trace can be evaluated with eq. (\ref{eqn:z0}) if we replace $g_1(L,l,\{S_i\})$ by $g^{\mathrm{min}}(L,l)$ or $g^{\mathrm{max}}(L,l)$. According to the intermediate value theorem, there is $g_1(L,l)$ such that $g_1^{\mathrm{min}}(L,l) \leq g_1(L,l)\leq g_1^{\mathrm{max}}(L,l)$, using which eq. (\ref{eqn:PF}) holds.
Furthermore, since both of $g_1^{\mathrm{min}}(L,l)$ and $g_1^{\mathrm{max}}(L,l)$ tend to zero in the limit
$L\rightarrow\infty, l\rightarrow\infty$, $g_1(L,l)$ also tends to zero.

Now, the law of large numbers is expressed as 
\begin{equation}
\sum_{i\in B_p}\ln{Z_0\{\beta(J\ \overline{m}_q+h_i)}\}=l^d [\ln{Z_0\{\beta(J\ \overline{m}_q+h)}\}]+l^dg_2(l,\{S_i\}),\label{eqn:lln}
\end{equation}
where $g_2(l,\{S_i\})$ converges to zero in the limit $l\rightarrow \infty$ as long as the variance of the stochastic variable on the left-hand side is finite. The law of large numbers originally means 
\begin{equation}
\mathrm{Prob}\left\{\left|\sum_{i\in B_p}\ln{Z_0\{\beta(J\, \overline{m}_q+h_i)}\}/l^d- [\ln{Z_0\{\beta(J\ \overline{m}_q+h_i)}\}]\right|\right\}>\epsilon\}\rightarrow0~(\forall\epsilon>0,~l^d\rightarrow\infty),\label{eqn:lnl}
\end{equation}
where $\mathrm{Prob}\{\cdot\}$ is the probability of the condition $(\cdot)$ to occur. Hence, when $l^d\rightarrow\infty$, only the situation where the absolute value in eq. (\ref{eqn:lnl}) is equal to zero can occur. This means that the argument on the left-hand side of eq. (\ref{eqn:lnl}) tends to $0$ in the limit $N\rightarrow\infty$. This is equivalent to eq. (\ref{eqn:lln}). 
This property can be applied to the partition function (\ref{eqn:PF}), leading to
\begin{multline}
Z(\beta,m)=\left(\frac{\beta J}{2\pi}\right)^\Omega\idotsint_{\mathbb{R}^\Omega \bigotimes (\mathit{i}\mathbb{R})^\Omega} \prod_{p}\mathit{d}m_p\mathit{d}\overline{m}_p\delta\left(\frac{\sum_{t=1}^{\Omega}m_t}{\Omega}-m\right)\\
		\exp{\left( \frac{\beta J}{2}\sum_{a,r} { U_{qr} m_q m_r}-\beta J\sum_{q}l^d \overline{m}_qm_q+\sum_{q}{l^d\left[\ln{Z_0\{\beta(J\ \overline{m}_q+h)}\}\right]}+\beta L^d g(L,l)\right)},\label{eqn:PF3}
\end{multline}
where the function $g(L,l)$ is defined as
\begin{equation}
\beta g(L,l)=\beta g_1(L,l)+g_2(l).
\end{equation}
For the same reason as before, the dependence of $g_2$ on $S_i$ can be ignored.  With this definition, the function $g(L,l)$ converges to zero in the limit $L\rightarrow \infty, l\rightarrow \infty$ regardless of the state of spins $S_i$ and the distribution of random fields $h_i$.

Next, we take the continuum limit of space in eq. (\ref{eqn:PF3}),
\begin{equation}
Z(\beta,m)=\left(\frac{\beta J}{2\pi}\right)^\Omega\iint\,{{\mathcal{D}}m(\vect{x}){\mathcal{D}}\overline{m}(\vect{x})}\delta\left(\int_{\mathbb{I}^d}\mathit{d}^dx\,m(\vect{x})-m\right){\exp\left(-\beta L^d{\mathcal{F}}\{\beta;m(\vect{x}),\overline{m}(\vect{x})\}+\beta L^dg(L,l)\right)},
\end{equation}
where 
\begin{multline}
{\mathcal{F}}\{\beta;m(\vect{x}),\overline{m}(\vect{x})\}=-\frac{J}{2}\iint_{{\mathbb{I}}^d \bigotimes {\mathbb{I}}^d}{\mathit{d}^dx\, \mathit{d}^dy\, U(\vect{x}-\vect{y})m(\vect{x})m(\vect{y})}\\
		+J\int_{{\mathbb{I}}^d}{\mathit{d}^dx\,\overline{m}(\vect{x})m(\vect{x})}-\frac{1}{\beta}\int_{{\mathbb{I}}^d}{\mathit{d}^dx\left[\ln{Z_0\{\beta(J\,\overline{m}(\vect{x})+h)}\}\right]},
\end{multline}
with ${\mathcal{D}}m(\vect{x})=\lim_{\Omega\rightarrow\infty}\Pi_{p=1}^{\Omega}dm_p$.  
When the saddle-point method is applied to $\overline{m}(\vect{x})$, the saddle-point equation is given as 
\begin{equation}
m(\vect{x})=\mathcal{T}_{\beta}\{\overline{m}(\vect{x})\}.
\end{equation}
The function $\mathcal{T}_{\beta}(x)$ is an increasing function because the definition leads to 
\begin{equation}
\frac{\partial}{\partial x}\mathcal{T}_{\beta}(x)=J\beta\left[\frac{\mathrm{tr}S^2\mathit{e}^{\beta(Jx+h)S}}{\mathrm{tr}\,\mathit{e}^{\beta(Jx+h)S}}-\left\{\frac{\mathrm{tr}S\mathit{e}^{\beta(Jx+h)S}}{\mathrm{tr}\,\mathit{e}^{\beta(Jx+h)S}}\right\}^2\right]\geq0.
\end{equation}
Hence the inverse function of $\mathcal{T}_{\beta}(x)$ can be defined. With the inverse function $\mathcal{T}_{\beta}^{-1}\{m(\vect{x})\}$, the generalized free energy ${\mathcal{F}}\{\beta;m(\vect{x})\}$ is expressed as
\begin{multline}
{\mathcal{F}}\{\beta;m(\vect{x})\}=-\frac{J}{2}\iint_{{\mathbb{I}}^d \bigotimes {\mathbb{I}}^d}{\mathit{d}^dx\, \mathit{d}^dy\, U(\vect{x}-\vect{y})m(\vect{x})m(\vect{y})}\\
+J\int_{{\mathbb{I}}^d}{\mathit{d}^dx\,\mathcal{T}_{\beta}^{-1}\{m(\vect{x})\}m(\vect{x})}-\frac{1}{\beta}\int_{{\mathbb{I}}^d}{\mathit{d}^dx\left[\ln{Z_0\{\beta(J\,\mathcal{T}_{\beta}^{-1}\{m(\vect{x})\}+h)}\}\right]}.
\end{multline}
By applying the saddle point method to $m(\vect{x})$, we evaluate the partition function and then obtain the free energy per spin as
\begin{align}
f(\beta,m)=\lim_{L,l,L/l \rightarrow \infty}{\left(-\frac{1}{\beta L^d}\ln{Z(\beta,m)}\right)}=\inf_{m(\svect{x})}\left\{\mathcal{F}\{\beta;m(\vect{x})\},\int_{\mathbb{I}^d}\mathit{d}^dx\,m(\vect{x})=m\right\}\label{eqn:f},
\end{align}
where $\inf\{f,\ast\}$ means the lower limit of $f$ under the condition $(\ast)$.

\section{Evaluation of the Generalized Free Energy and Its Derivative}\label{app:GFE}
In this Appendix, we derive eqs. (\ref{eqn:1}) and (\ref{eqn:2d}).

First, eq.  (\ref{eqn:1}) is evaluated as follows.
\begin{align}
{\mathcal{F}}\{\beta;m(\vect{x})\}&\geq-\frac{J}{2}m^2-\frac{J}{2}U_{\mathrm{max}}\left\{\int_{{\mathbb{I}}^d}\mathit{d}^dx\, m(\vect{x})^2-m^2\right\}\notag\\
&+J\int_{{\mathbb{I}}^d}{\mathit{d}^dx\,\mathcal{T}_{\beta}^{-1}\{m(\vect{x})\}m(\vect{x})}-\frac{1}{\beta}\int_{{\mathbb{I}}^d}{\mathit{d}^dx\left[\ln{Z_0\{\beta(J\,\mathcal{T}_{\beta}^{-1}\{m(\vect{x})\}+h)}\}\right]}\notag\\
&=-\frac{J}{2}m^2+J\mathcal{T}_{\beta}^{-1}\{m\}m-\frac{1}{\beta}\left[\ln{Z_0\{\beta(J\,\mathcal{T}_{\beta}^{-1}\{m\}+h)}\}\right]\notag\\
&-\frac{J}{2}U_{\mathrm{max}}\int_{{\mathbb{I}}^d}\mathit{d}^dx\, m(\vect{x})^2+J\int_{{\mathbb{I}}^d}{\mathit{d}^dx\,\mathcal{T}_{\beta}^{-1}\{m(\vect{x})\}m(\vect{x})}-\frac{1}{\beta}\int_{{\mathbb{I}}^d}{\mathit{d}^dx\left[\ln{Z_0\{\beta(J\,\mathcal{T}_{\beta}^{-1}\{m(\vect{x})\}+h)}\}\right]}\notag\\
&+\frac{J}{2}U_{\mathrm{max}}m^2-J\mathcal{T}_{\beta}^{-1}\{m\}m+\frac{1}{\beta}\left[\ln{Z_0\{\beta(J\,\mathcal{T}_{\beta}^{-1}\{m\}+h)}\}\right]\notag\\
&=f_{\mathrm{MF}}\{\beta,m\}\notag\\
&+U_{\mathrm{max}}\int_{{\mathbb{I}}^d}\mathit{d}^dx\,\left\{-\frac{J}{2}m(\vect{x})^2+J\frac{\mathcal{T}_{\beta}^{-1}\{m(\vect{x})\}}{U_{\mathrm{max}}}m(\vect{x})-\frac{1}{\beta U_{\mathrm{max}}}\left[\ln{Z_0\left\{\beta U_{\mathrm{max}}\left(J\,\frac{\mathcal{T}_{\beta}^{-1}\{m(\vect{x})\}}{U_{\mathrm{max}}}+\frac{h}{U_{\mathrm{max}}}\right)\right\}}\right]\right\}\notag\\
&-U_{\mathrm{max}}\left\{-\frac{J}{2}m^2+J\frac{\mathcal{T}_{\beta}^{-1}\{m\}}{U_{\mathrm{max}}}m-\frac{1}{\beta U_{\mathrm{max}}}\left[\ln{Z_0\left\{\beta U_{\mathrm{max}}\left(J\,\frac{\mathcal{T}_{\beta}^{-1}\{m\}}{U_{\mathrm{max}}}+\frac{h}{U_{\mathrm{max}}}\right)\right\}}\right]\right\}\notag\\
&=f_{\mathrm{MF}}\{\beta,m\}+U_{\mathrm{max}}\left\{\int_{{\mathbb{I}}^d}\mathit{d}^dx\,f_{\mathrm{MF}}^{*}\{\beta^{*},m(\vect{x})\}-f_{\mathrm{MF}}^{*}\{\beta^{*},m\}\right\}\label{eqn:1A}.
\end{align}
Equation (\ref{eqn:2d}) is derived as
\begin{align}
\left.\frac{\partial^2\mathcal{F}\{\beta\,;\,m(\vect{x})\}}{\partial m_{-\svect{n}'}\partial m_{\svect{n}}}\right|_{m(\svect{x})=m}
=&\frac{\partial}{\partial m_{-\svect{n}'}}
\left\{JU_{\svect{n}}m_{-\svect{n}}
+J\int_{{\mathbb{I}}^d}{\mathit{d}^dx\,\frac{\partial \mathcal{T}_{\beta}^{-1}\{m(\vect{x})\}}{\partial m(\vect{x})}\mathit{e}^{-2\pi i\svect{n}\cdot\svect{x}}m(\vect{x})}\right.\notag\\&
+J\int_{{\mathbb{I}}^d}{\mathit{d}^dx\,\mathcal{T}_{\beta}^{-1}\{m(\vect{x})\}\mathit{e}^{-2\pi i\svect{n}\cdot\svect{x}}}\notag\\&
\left.\left.-\frac{1}{\beta}\int_{{\mathbb{I}}^d}{\mathit{d}^dx\,\frac{\partial}{\partial \mathcal{T}_{\beta}^{-1}\{m(\vect{x})\}}\left[\ln{Z_0\{\beta(J\,\mathcal{T}_{\beta}^{-1}\{m(\vect{x})\}+h)}\}\right]}\frac{\partial \mathcal{T}_{\beta}^{-1}\{m(\vect{x})\}}{\partial m(\vect{x})}\mathit{e}^{-2\pi i\svect{n}\cdot\svect{x}}
\right\}\right|_{m(\svect{x})=m}\notag\\
=&\frac{\partial}{\partial m_{-\svect{n}'}}
\left.\left\{JU_{\svect{n}}m_{-\svect{n}}
+J\int_{{\mathbb{I}}^d}{\mathit{d}^dx\,\mathcal{T}_{\beta}^{-1}\{m(\vect{x})\}\mathit{e}^{-2\pi i\svect{n}\cdot\svect{x}}}\right\}\right|_{m(\svect{x})=m}\notag\\
=&\left.-JU_{\svect{n}}\delta_{\svect{n}'\svect{n}}+J\int_{{\mathbb{I}}^d}{\mathit{d}^dx\,\frac{\partial \mathcal{T}_{\beta}^{-1}\{m(\vect{x})\}}{\partial m(\vect{x})}\mathit{e}^{-2\pi i(\svect{n}-\svect{n}')\cdot\svect{x}}}\right|_{m(\svect{x})=m}\notag\\
=&-JU_{\svect{n}}\delta_{\svect{n}'\svect{n}}+J\frac{\partial \mathcal{T}_{\beta}^{-1}\{m\}}{\partial m}\int_{{\mathbb{I}}^d}\mathit{d}^dx\,\mathit{e}^{-2\pi i(\svect{n}-\svect{n}')\cdot\svect{x}}\notag\\
=&\left(-JU_{\svect{n}}+J\frac{\partial \mathcal{T}_{\beta}^{-1}\{m\}}{\partial m}\right)\delta_{\svect{n}'\svect{n}}\label{eqn:2dA}
\end{align}

\section{Normalized Potential in the van dar Waals Limit}\label{app:vdW}
In this section, we prove that the potential $U(\vect{x})$ in the van dar Waals limit is the delta function $\delta(\vect{x})$ by showing that $(\gamma L)^d\phi(\gamma L\vect{x})$ approaches the delta function.

Let the $\epsilon$-vicinity of the origin be written as $B_\epsilon=\{\vect{x}\in \mathbb{I}^d :|\vect{x}|<\epsilon\}$ and $\mathbb{I}^d\setminus B_\epsilon$ be denoted as $\overline{B}_\epsilon$.
The integral of $(\gamma L)^d\phi(\gamma L\vect{x})$ over $\overline{B}_\epsilon$ is
\begin{align}
\int_{\overline{B}_\epsilon}\mathit{d}^dx\,(\gamma L)^d\phi(\gamma L\vect{x})=\int_{\gamma L\overline{B}_\epsilon}\mathit{d}^dx'\,\phi(\vect{x}'),\label{eqn:Inte}
\end{align}
where $\gamma L\overline{B}_\epsilon=\{\vect{x}\in (\gamma L)^d \mathbb{I}^d:|\vect{x}|\ge\gamma L \epsilon\}$.
In the limit $L\rightarrow\infty$, all points in $\gamma L\overline{B}_\epsilon$ tend to $\infty$.
Hence, for the integrable function $\psi(\vect{x})$, the integral (\ref{eqn:Inte}) tends to $0$,
\begin{align}
\lim_{\gamma\rightarrow0}\lim_{L\rightarrow\infty}\int_{\gamma L\overline{B}_\epsilon}\mathit{d}^dx'\,\phi(\vect{x}')=0.\label{eqn:Inte2}
\end{align}
On the other hand, the integral over $\mathbb{I}^d$ is divided into two parts,
\begin{equation}
1=\int_{\mathbb{I}^d}\mathit{d}^dx\,(\gamma L)^d\phi(\gamma L\vect{x})=\int_{B_\epsilon}\mathit{d}^dx\,(\gamma L)^d\phi(\gamma L\vect{x})+\int_{\overline{B}_\epsilon}\mathit{d}^dx\,(\gamma L)^d\phi(\gamma L\vect{x}),\label{eqn:Inte3}
\end{equation}
where we have used eq. (\ref{eqn:normU}).
In the van der Waals limit, the above equation implies
\begin{align}
\lim_{\gamma\rightarrow0}\lim_{L\rightarrow\infty}\int_{\mathbb{I}^d}\mathit{d}^dx\,(\gamma L)^d\phi(\gamma L\vect{x})&=\lim_{\gamma\rightarrow0}\lim_{L\rightarrow\infty}\int_{B_\epsilon}\mathit{d}^dx\,(\gamma L)^d\phi(\gamma L\vect{x})+\lim_{\gamma\rightarrow0}\lim_{L\rightarrow\infty}\int_{\overline{B}_\epsilon}\mathit{d}^dx\,(\gamma L)^d\phi(\gamma L\vect{x})\notag\\
&=\lim_{\gamma\rightarrow0}\lim_{L\rightarrow\infty}\int_{B_\epsilon}\mathit{d}^dx\,(\gamma L)^d\phi(\gamma L\vect{x})=1. \label{apb4}
\end{align}

Next, let us prepare a test function $f(\vect{x})$ that is continuous and integrable, and evaluate the integral of $f(\vect{x})(\gamma L)^d\phi(\gamma L\vect{x})$. If we can prove the following equation,
\begin{align}
\lim_{\gamma\rightarrow0}\lim_{L\rightarrow\infty}\int_{\mathbb{I}^d}\mathit{d}^dx\,f(\vect{x})(\gamma L)^d\phi(\gamma L\vect{x})=f(\vect{0}),\label{eqn:fin}
\end{align}
$(\gamma L)^d\phi(\gamma L\vect{x})$ is confirmed to approach the delta function. 
The integral of $f(\vect{x})(\gamma L)^d\phi(\gamma L\vect{x})$ is divided in two parts as in eq. (\ref{eqn:Inte3}),
\begin{equation}
\int_{\mathbb{I}^d}\mathit{d}^dx\,f(\vect{x})(\gamma L)^d\,\phi(\gamma L\vect{x})=\int_{B_\epsilon}\mathit{d}^dx\,f(\vect{x})(\gamma L)^d\phi(\gamma L\vect{x})+\int_{\overline{B}_\epsilon}\mathit{d}^dx\,f(\vect{x})(\gamma L)^d\phi(\gamma L\vect{x}).\label{eqn:Inte4}
\end{equation}
The first term on the right-hand side of eq. (\ref{eqn:Inte4}) is bounded as
\begin{align}
\Big( \inf_{\svect{x}'\in B_\epsilon}f(\vect{x}')\Big)\cdot
\int_{B_\epsilon}\mathit{d}^dx\,(\gamma L)^d\phi(\gamma L\vect{x})\le\int_{B_\epsilon}\mathit{d}^dx\,f(\vect{x})(\gamma L)^d\phi(\gamma L\vect{x})\le \Big(\sup_{\svect{x}'\in B_\epsilon}f(\vect{x}')\Big)\cdot
\int_{B_\epsilon}\mathit{d}^dx\,(\gamma L)^d\phi(\gamma L\vect{x}).
\end{align}
In the van der Waals limit, the above equation is reduced to, according to eq. (\ref{apb4}).
\begin{align}
\inf_{\svect{x}'\in B_\epsilon}f(\vect{x}')\le\lim_{\gamma\rightarrow0}\lim_{L\rightarrow\infty}\int_{B_\epsilon}\mathit{d}^dx\,f(\vect{x})(\gamma L)^d\phi(\gamma L\vect{x})\le\sup_{\svect{x}'\in B_\epsilon}f(\vect{x}')\label{eqn:Inte5}.
\end{align}
The second term on the right-hand side of eq. (\ref{eqn:Inte4})  is evaluated as
\begin{align}
\inf_{\svect{x}'\in \overline{B}_\epsilon}\{f(\vect{x}')-f(\vect{0})\}\int_{\overline{B}_\epsilon}\mathit{d}^dx\,(\gamma L)^d\phi(\gamma L\vect{x})\le&\int_{\overline{B}_\epsilon}\mathit{d}^dx\,\{f(\vect{x})-f(\vect{0})\}(\gamma L)^d\phi(\gamma L\vect{x})\notag\\
\le&\sup_{\svect{x}'\in \overline{B}_\epsilon}\{f(\vect{x}')-f(\vect{0})\}\int_{\overline{B}_\epsilon}\mathit{d}^dx\,(\gamma L)^d\phi(\gamma L\vect{x}).
\end{align}
In the van der Waals limit, the right- and left-hand sides both tend to $0$ due to eq. (\ref{eqn:Inte2}), which leads to 
\begin{align}
\lim_{\gamma\rightarrow0}\lim_{L\rightarrow\infty}\int_{\overline{B}_\epsilon}\mathit{d}^dx\,\{f(\vect{x})-f(\vect{0})\}(\gamma L)^d\phi(\gamma L\vect{x})=0.\label{eqn:Inte6}
\end{align}
With eqs. (\ref{eqn:Inte4}), (\ref{eqn:Inte5}), and (\ref{eqn:Inte6}), we find
\begin{align}
\inf_{\svect{x}'\in B_\epsilon}f(\vect{x}')\le\lim_{\gamma\rightarrow0}\lim_{L\rightarrow\infty}\int_{\mathbb{I}^d}\mathit{d}^dx\,f(\vect{x})(\gamma L)^d\,\phi(\gamma L\vect{x})\le\sup_{\svect{x}'\in B_\epsilon}f(\vect{x}').
\end{align}
The above inequality is reduced to
\begin{align}
\left|\lim_{\gamma\rightarrow0}\lim_{L\rightarrow\infty}\int_{\mathbb{I}^d}\mathit{d}^dx\,f(\vect{x})(\gamma L)^d\,\phi(\gamma L\vect{x})-f(\vect{0})\right|&\le
\max\left\{\left|\inf_{\svect{x}'\in B_\epsilon}f(\vect{x}')-f(\vect{0})\right|,\left|\sup_{\svect{x}'\in B_\epsilon}f(\vect{x}')-f(\vect{0})\right|\right\}\notag\\
&\le\sup_{\svect{x}'\in B_\epsilon}\big|f(\vect{x}')-f(\vect{0})\big|.\label{eqn:Inte7}
\end{align}
Next, let us remember that the function $f(\vect{x})$ is continuous:
\begin{align}
\sup_{\svect{x}\in B_{\delta}}|f(\vect{x})-f(\vect{0})|<\epsilon'~(\forall\epsilon', \exists\delta>0).\label{eqn:Inte8}
\end{align}
With eqs. (\ref{eqn:Inte7}) and (\ref{eqn:Inte8}), we can state there is $\delta$ such that 
\begin{align}
\left|\lim_{\gamma\rightarrow0}\lim_{L\rightarrow\infty}\int_{\mathbb{I}^d}\mathit{d}^dx\,f(\vect{x})(\gamma L)^d\,\phi(\gamma L\vect{x})-f(\vect{0})\right|<\epsilon~(\forall\epsilon>0),
\end{align}
which means eq. (\ref{eqn:fin}).

\end{document}